\documentclass[useAMS,usenatbib]{mn2e}

\usepackage{graphicx}
\usepackage{amssymb}
\usepackage{amsmath}

\newcommand{\diff}{\ensuremath{\rm d}}

\newcommand{\va}{\ensuremath{\mathbf{a}}}
\newcommand{\vzero}{\ensuremath{\mathbf{0}}}

\newcommand{\Rpole}{\ensuremath{R_{\rm p}}}
\newcommand{\Rstar}{\ensuremath{R_{\ast}}}
\newcommand{\Rkep}{\ensuremath{R_{\rm K}}}
\newcommand{\epsstar}{\ensuremath{\epsilon_{\ast}}}
\newcommand{\Mdot}{\ensuremath{\dot{M}}}

\newcommand{\Omegacrit}{\ensuremath{\Omega_{\rm c}}}

\newcommand{\tauz}{\ensuremath{\tau_{0}}}

\newcommand{\flux}{\ensuremath{F}}
\newcommand{\fluxstar}{\ensuremath{\flux_{\ast}}}
\newcommand{\dflux}{\ensuremath{\Delta\flux}}

\newcommand{\phase}{\ensuremath{\phi}}
\newcommand{\dphase}{\ensuremath{\Delta \phase}}

\newcommand{\ibeta}{\ensuremath{\{i,\beta\}}}

\newcommand{\column}{\ensuremath{D}}

\newcommand{\dpot}{\ensuremath{\Psi}}

\newcommand{\hm}{\ensuremath{h_{\rm m}}}

\newcommand{\drmag}{\ensuremath{\gamma}}
\newcommand{\pmag}{\ensuremath{\tilde{\phi}}}
\newcommand{\tmag}{\ensuremath{\tilde{\theta}}}

\newcommand{\xobs}{\ensuremath{x_{\rm o}}}
\newcommand{\yobs}{\ensuremath{y_{\rm o}}}
\newcommand{\zobs}{\ensuremath{z_{\rm o}}}

\newcommand{\Rtor}{\ensuremath{R_{\rm t}}}
\newcommand{\dtor}{\ensuremath{d_{\rm t}}}

\newcommand{\arclen}{\ensuremath{\ell}}
\newcommand{\sma}{\ensuremath{b}}

\newcommand{\ang}{\ensuremath{\upsilon}}
\newcommand{\angtor}{\ensuremath{\ang_{\rm t}}}
\newcommand{\angmag}{\ensuremath{\tilde{\ang}}}

\newcommand{\epsp}{\ensuremath{\tilde{\delta}_{\phi}}}
\newcommand{\epst}{\ensuremath{\tilde{\delta}_{\theta}}}

\newcommand{\sorie}{$\sigma$~Ori~E}

\newcommand{\Halpha}{H$\alpha$}

\title[Magnetospheres around Helium-strong stars]
      {Exploring the photometric signatures of magnetospheres
       around Helium-strong stars}
\author[R. H. D. Townsend]
       {R. H. D. Townsend\thanks{E-mail: rhdt@bartol.udel.edu}\\
Bartol Research Institute,
Department of Physics \& Astronomy,
University of Delaware,
Newark, DE 19716, USA}

\date{%
Received: .................................... 
Accepted: ....................................
}

\pagerange{\pageref{firstpage}--\pageref{lastpage}}
\pubyear{2007}

\begin{document}


\maketitle

\label{firstpage}

\begin{abstract}
The photometric variations due to magnetically confined material
around He-strong stars are investigated within the framework of the
Rigidly Rotating Magnetosphere (RRM) model. For dipole field
topologies, the model is used to explore how the morphology of light
curves evolves in response to changes to the observer inclination,
magnetic obliquity, rotation rate and optical depth.

The general result is that double-minimum light curves arise when the
obliquity and/or inclination are close to $90\degr$; no light
variations are seen in the opposite limit; and for intermediate cases,
single-minimum light curves occur. These findings are interpreted with
the aid of a simple, analytical torus model, paving the way for the
development of new photometric-based constraints on the fundamental
parameters of He-strong stars. Illustrative applications to five stars
in the class are presented.
\end{abstract}



\section{Introduction} \label{sec:intro}

The Helium-strong stars are small class of chemically peculiar B
stars, interpreted as an extension of the Ap/Bp phenomenon to higher
effective temperatures \citep{OsmPet1974}. They are characterized by
elevated and spatially inhomogeneous photospheric Helium abundances,
and in most cases also exhibit strong magnetic fields \citep[see][and
references therein]{Boh1987}. Spectroscopy of their time-varying
Helium absorption lines often reveals that the surface abundance
distribution is strongly correlated with field
topology \citep{BohLan1988,Vet1990}. These correlations lend support
to models for the abundance anomalies that rely on the interplay
between radiative and magnetic forces, either in the framework of
elemental diffusion \citep[e.g.,][]{Mic1970}, or in the context of
wind mass loss \citep[e.g.,][]{HunGro1999}.

Of the He-strong stars exhibiting photometric variability
\citep{PedTho1977}, at least some can successfully be explained by
invoking rotational modulation of the visible abundance
inhomogeneities -- so-called `spot' models \citep[see,
e.g.,][]{Krt2007}. However, a different approach is required for other
objects in the class, most notably the B2pe star \sorie. The
light curve of this star has a double-eclipse morphology
\citep*{Hes1977}, leading some authors \citep*[e.g.,][]{Hes1976}
initially to suggest that the star is a mass-transfer binary. In
fact, as further observational data have become available, an
empirical picture for the star's photometric and spectroscopic
variability has emerged \citep{GroHun1982}, that augments a spot model
with a pair of circumstellar clouds. These clouds absorb continuum
photons as they transit the stellar disk, giving rise to the
distinctive photometric minima; however, once off the disk they reveal
themselves in \Halpha\ line profiles as blue- and red-shifted emission
components \citep[e.g.,][]{WalHes1976,Ped1979}.

Recently, \citet[][hereafter TO-05]{TowOwo2005} have developed a
Rigidly Rotating Magnetosphere (RRM) model that provides the
theoretical basis for this empirical picture. The model conjectures
that plasma in the star's radiatively driven wind, channeled by nearly
rigid field lines, tends to accumulate at points where the effective
(gravitational plus centrifugal) potential is at a local minimum. This
steady accumulation leads to the development of a magnetosphere
resembling a warped disk, which is supported by the centrifugal force
and co-rotates rigidly with the star. For an oblique dipole field
topology, the RRM model predicts that the magnetospheric density is
highest in two regions -- `clouds' -- situated along the twin
intersections between the magnetic and rotational equators. In the
specific case of \sorie, the photometric and \Halpha\ variations
synthesized using the model show very good agreement with observations
of the star \citep*{Tow2005}.

In the present paper I apply the RRM model in a broader context, by
exploring how the photometric signatures of magnetospheric plasma
depend on fundamental parameters such as the observer inclination and
magnetic obliquity (\S\ref{sec:explore}). The results from this
parameter-space exploration are interpreted with the aid of a simple,
analytic torus model (\S\ref{sec:torus}), and I discuss how they might
be applied to constrain the parameters of He-strong stars
(\S\ref{sec:discuss}). The principal findings are then summarized
(\S\ref{sec:summary}).

\section{Exploration of parameter space} \label{sec:explore}

\subsection{Light-curve synthesis} \label{ssec:explore-light}

The RRM model as applied to a dipole magnetic topology is used to
synthesize the light curves used for exploring parameter space. The
same procedure described by \citet[][their \S3.2]{Tow2005} is
followed; note therefore that the variability arising from abundance
inhomogeneities is not considered. It is assumed that there is no
decentering of the magnetic field, $\va = \vzero$ (eqn. 1, ibid.), and
a canonical energy ratio parameter $\epsstar = 10^{-3}$ (cf. TO-05,
eqn. 38) is adopted throughout.

The remaining free parameters of the model are then the inclination
 $i$; the magnetic obliquity $\beta$; the ratio
\begin{equation}
\omega = \frac{\Omega}{\Omegacrit}
\end{equation}
between the rotation frequency $\Omega$ and the critical frequency
 $\Omegacrit$; and an optical depth scale
\begin{equation}
\tauz = \frac{\kappa\, \Mdot\, t\, \omega^{4/3}}{\Rstar^{2}}.
\end{equation}
This latter expression groups the opacity $\kappa$ (assumed constant)
with various terms appearing in the RRM prescription for the density
(TO-05, eqn. 35). The additional $\omega^{4/3}$ term comes from
considering the rotational dependence of the total magnetospheric mass
column between star and observer; its inclusion ensures that models
having the same \tauz\ but differing $\omega$ exhibit approximately
the same degree of light variations.

\subsection{Varying $\beta$ and $i$} \label{ssec:explore-beta-i}

\begin{figure*}
\begin{center}
\includegraphics{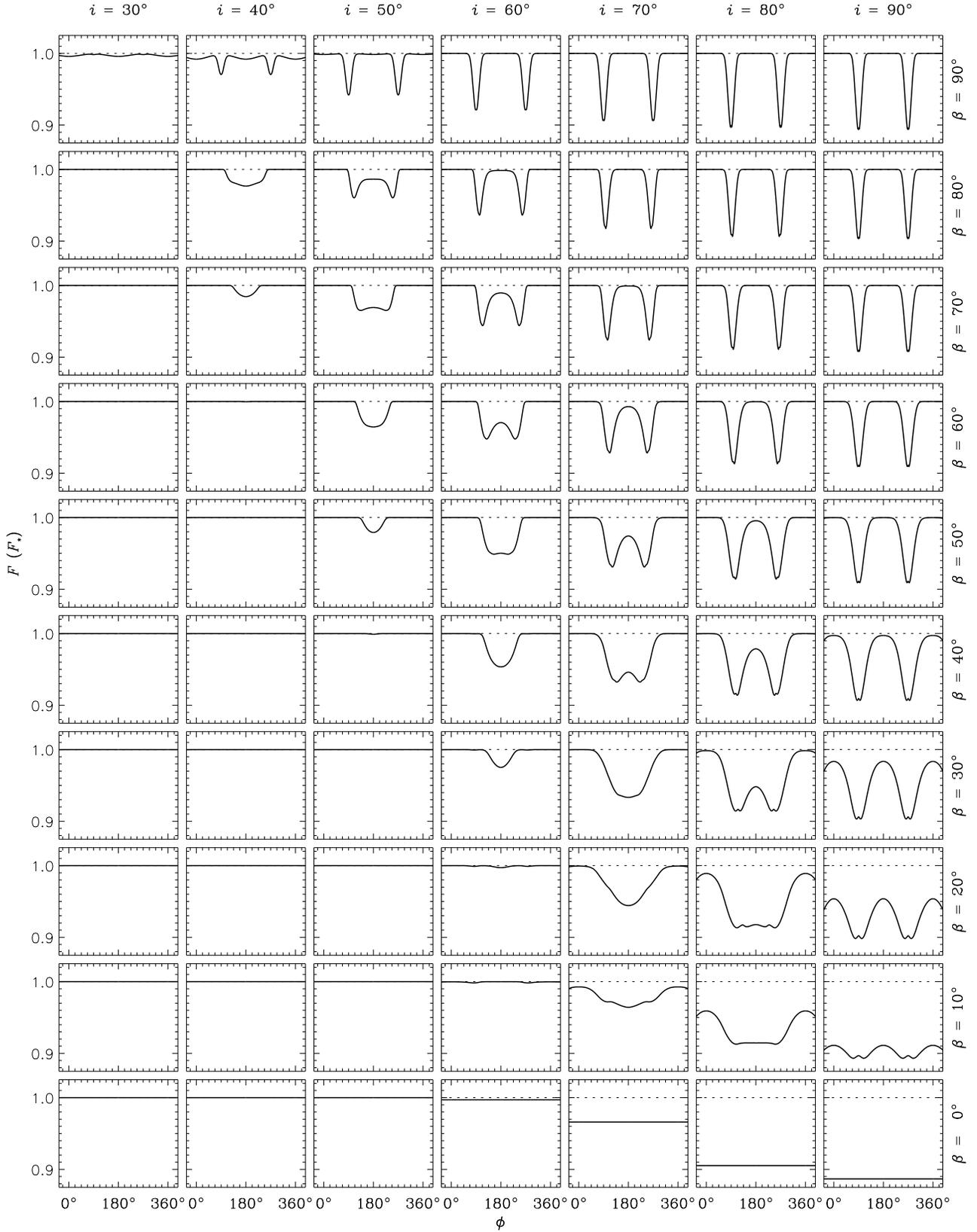}
\end{center}
\caption{Light curves synthesized with varying inclination $i$ and
obliquity $\beta$, in each case plotting the flux \flux\ (in units of
the unobscured stellar flux \fluxstar) as a function of the rotation
phase \phase; a rotation rate $\omega = 0.5$ and an optical depth
scale $\tauz = 10$ are assumed. The dotted horizontal lines indicate
the level $\flux = \fluxstar$.}
\label{fig:beta-i-curves}
\end{figure*}

To begin the exploration of parameter space, a grid of light curves is
synthesized in $5\degr$ increments over the full intervals $0\degr \le
i \le 90\degr$ and $0\degr \le \beta \le 90\degr$. In each case an
intermediate rotation rate of $\omega = 0.5$ is assumed, and an
optical depth scale $\tauz = 10$ is chosen to produce variability on
the $\sim 10\%$ level typical to He-strong
stars. Figure~\ref{fig:beta-i-curves} illustrates the resulting light
curves, plotting the flux \flux\ (in units of the unobscured stellar
flux \fluxstar) against the rotational phase \phase\footnote{In the
present work, $\phase=0\degr$ corresponds to magnetic maximum; this
differs from the convention adopted by \citet{Tow2005}, who followed
historical precedent \citep[e.g.,][]{Hes1977} by choosing the primary
light minimum as phase zero.}. For clarity only the curves at even
multiples of $5\degr$ are shown, and the $i < 30\degr$ cases are
omitted because none exhibit any significant variability.

The figure clearly demonstrates how the character of the light curves
is altered as $i$ and $\beta$ are changed. In the limit where either
of these parameters is large, the curves exhibit a clear
double-minimum (hereafter, `2-m') morphology. In the opposite limit,
and in particular for $i < 30\degr$ as already mentioned, a
non-varying, constant morphology (`0-m') is seen. For intermediate
cases, falling approximately along a band extending from $\ibeta =
\{30\degr,90\degr\}$ to $\ibeta = \{90\degr,0\degr\}$, a single light
minimum (`1-m') occurs at phase $\phase=180\degr$.

\begin{figure*}
\begin{center}
\includegraphics{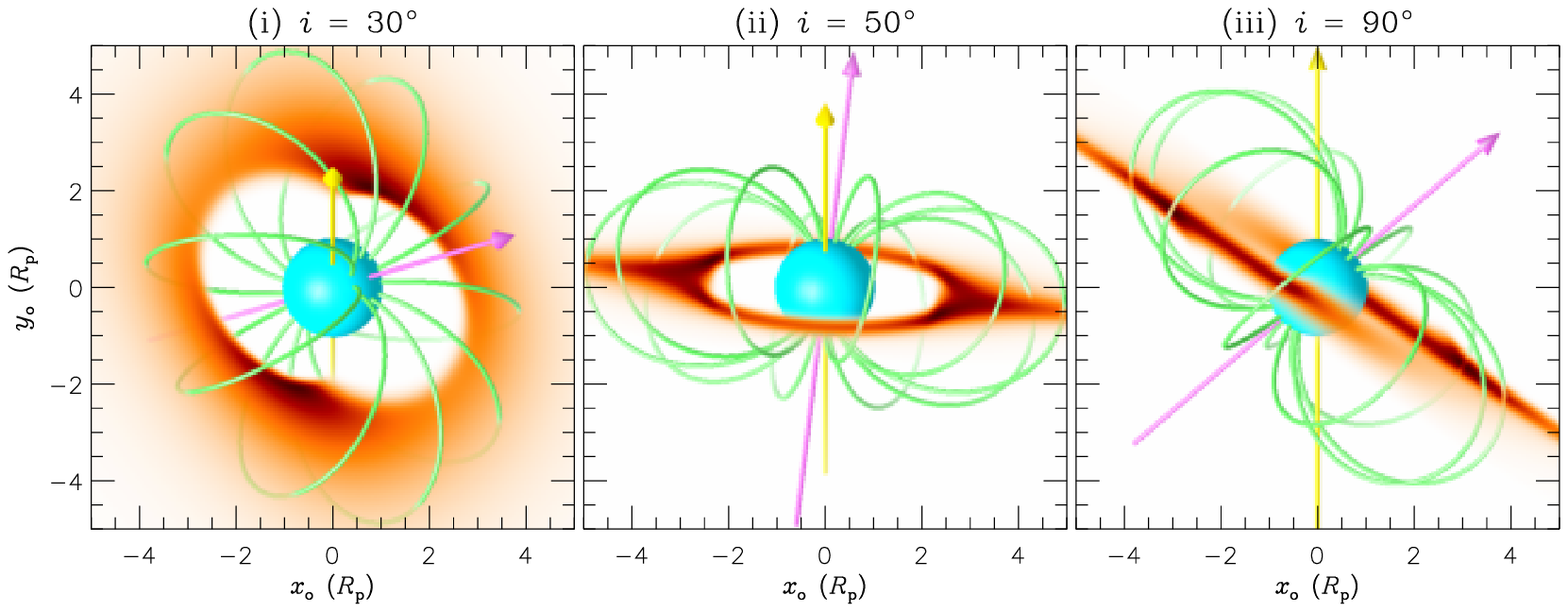}
\end{center}
\caption{Maps of the column density \column\ in the observer's image
  plane $(\xobs,\yobs)$, for three RRM configurations: (i) $\ibeta =
  \{30\degr,50\degr\}$ at a rotational phase $\phase = 81\degr$, (ii)
  $\ibeta = \{50\degr,50\degr\}$ at $\phase = 171\degr$, and (iii)
  $\ibeta = \{90\degr,50\degr\}$ at $\phase = 81\degr$. (The phases
  are offset slightly from the `round-number' values $90\degr$ and
  $180\degr$, to improve visual clues to the 3-D geometry). In all
  three cases, $\omega = 0.5$. Black/opaque corresponds to highest
  column density, and white/transparent to lowest, with intermediate
  levels shown in orange. The central star is coloured blue, and the
  axes are drawn as arrows --- yellow/upright for the rotation axis,
  and magenta/oblique for the magnetic axis. Moreover, selected field
  lines having a summit radius of $5\,\Rpole$ (where \Rpole\ is the
  stellar radius at the rotational poles) are indicated in green.}
\label{fig:maps}
\end{figure*}

To understand the origin of these three morphologies,
Fig.~\ref{fig:maps} shows maps of the column density \column\ in the
near-star regions of the circumstellar environment, for three
configurations having $\beta=50\degr$ and $i=30\degr$, $50\degr$, and
$90\degr$. The maps are calculated from an expression of the form
\begin{equation}
\column(\xobs,\yobs) = \int \rho(\xobs,\yobs,\zobs)\, \diff\zobs,
\end{equation}
where $(\xobs,\yobs,\zobs)$ are Cartesian coordinates in the
observer's reference frame, with \zobs\ directed toward the observer,
and $\rho$ is the local density predicted by the RRM model (cf. TO-05,
eqn.~35). Panel (i) of the figure, plotting the $i=30\degr$ case, is
typical of low-inclination systems.  With the observer situated close
to the rotation pole, the plasma trapped in the magnetosphere -- lying
in a disk between the rotational and magnetic equatorial planes -- has
no opportunity to pass between the observer and the star. Hence, an
unvarying 0-m light curve is seen in these systems.

For panel (iii) the converse is true: with the observer in the
rotational equatorial plane, the magnetospheric plasma transits the
star twice per rotation cycle, and thus a 2-m light curve arises. In
this particular case, symmetry requires that the minima occur at
phases $\phase=90\degr$ and $270\degr$, with a half-cycle difference
between them. However, as discussed in further detail below, a smaller
phase difference \dphase\ is found for 2-m configurations having $i <
90\degr$ and $\beta < 90\degr$.

Panel (ii) illustrates the intermediate, $i=50\degr$ case, where the
observer is not sufficiently close to the rotational equator for
magnetospheric plasma to pass fully in front of the star. Instead,
when the magnetic field is pointed away from the observer the plasma
grazes the lower portions of the stellar limb, causing a 1-m light
curve with the minimum centered at phase $\phase = 180\degr$. This
light curve is characterized by a reduced overall absorption, relative
to the 2-m case shown in panel (iii).

As can be seen from the $\beta=50\degr$ row of light curves in
Fig.~\ref{fig:beta-i-curves}, a smooth evolutionary sequence links the
cases shown in panels (i), (ii) and (iii) of Fig.~\ref{fig:maps}. As
the observer moves from the rotational equator toward smaller $i$, the
depth ($\equiv \fluxstar - \flux$) at the twin light minima of the
(initially, 2-m) light curve \emph{decreases}; whereas, the depth at
the intervening light maximum at $\phase=0.5$
\emph{increases}. Eventually, these depth values cross over, with the result
that the light curve transitions to a 1-m morphology. Following this
switch-over (which in the $\beta=50\degr$ case occurs at $i \approx
60\degr$), the depth at the single remaining minimum at
$\phase=180\degr$ declines toward lower $i$, and eventually vanishes
in a further transition to a 0-m morphology.

\begin{figure*}
\begin{center}
\includegraphics{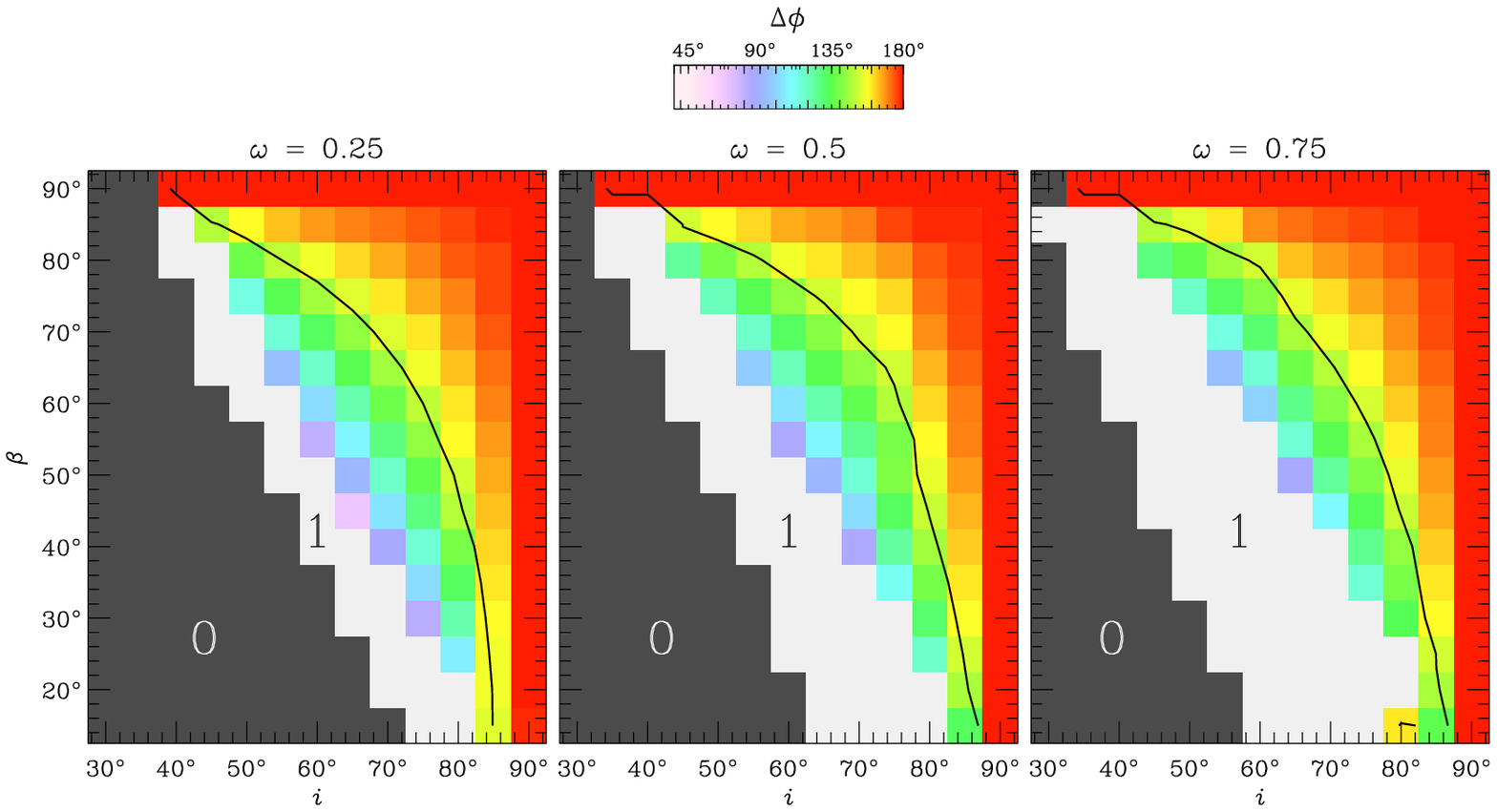}
\end{center}
\caption{The phase difference \dphase\ between the minima in 2-m light
curves, plotted in the $i$-$\beta$ plane for three differing rotation
rates; in every case, $\tauz = 10$. The curve in each panel shows the
contour $\dphase = 150\degr$, discussed in
\S\ref{sssec:discuss-app-sorie}. Regions of the plane associated with
0-m and 1-m curves are shown in monochrome and labeled as `0' and `1',
respectively.}
\label{fig:dphase}
\end{figure*}

The progressive distortion of light curves, as the 2-m to 1-m
transition is approached, has an influence on the phase difference
\dphase\ between the minima. To explore this influence, each light
curve in the $i$-$\beta$ grid is analyzed using a simple algorithm
that determines its morphology, and -- if 2-m -- measures the phase
difference. The algorithm involves three steps:

\begin{enumerate}
\item If the total flux variation $\dflux \equiv \max(\flux) -
  \min(\flux)$ is less than $0.01\,\fluxstar$, then the light curve is
  classified as 0-m.
\item Otherwise, if the light curve exhibits two flux minima,
  \emph{and} these minima are separated by flux maxima that both
  satisfy $\flux - \min(\flux) > 0.15 \dflux$, then the light curve is
  classified as 2-m.
\item Otherwise, the light curve is classified as 1-m.
\end{enumerate}

A minor complication arises when minima are cusped (see
\S\ref{ssec:explore-omega} and Fig.~\ref{fig:omega-curves}); then, the
adopted phase of a minimum is determined by averaging the phases of
the local minima on either side of the cusp.

Figure~\ref{fig:dphase} plots the results of this analysis, showing
the \dphase\ values for the 2-m light curves in the $i$-$\beta$ plane,
and also indicating where in the plane 1-m and 0-m curves occur. (The
middle panel of the figure shows the $\omega=0.5$ case considered in
this section; the other two panels, corresponding to
$\omega=0.25$ and $\omega=0.75$, are discussed in the following
section). As before, the $i < 30\degr$ region of the plane is
uniformly 0-m and is therefore not shown; but in the present case, the
$\beta < 20\degr$ region is also omitted, because there the analysis
algorithm gives unreliable results. This is more due to the difficulty
in sensibly classifying the light curves in this region (in
particular, see the $\ibeta=\{80\degr,10\degr\}$ and
$\{90\degr,10\degr\}$ cases shown in Fig.~\ref{fig:beta-i-curves}),
than to any particular limitation of the algorithm.

The figure confirms the previous finding that the 0-m and 2-m light
curves are situated toward small and large angles, respectively, with
the 1-m curves falling in between. However, it also reveals that
\dphase\ depends quite regularly on $i$ and $\beta$. When either of
these parameters is equal to $90\degr$, the phase difference is
exactly one-half of a cycle; as already mentioned, this is required by
symmetry. Away from these special cases, and toward the boundary
between the 2-m and 1-m regions, \dphase\ is constant along contours
running from low-$i$/high-$\beta$ corner down to the
high-$i$/low-$\beta$ corner. The origin of this behaviour is explored
further in \S\ref{sec:torus}.

\subsection{Varying $\omega$} \label{ssec:explore-omega}

\begin{figure}
\begin{center}
\includegraphics{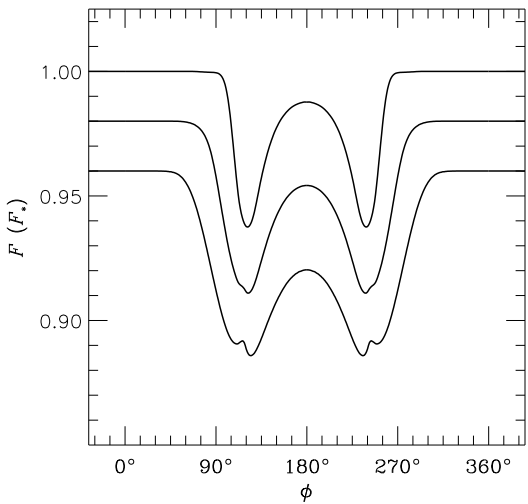}
\end{center}
\caption{Light curves for a $\ibeta = \{70\degr,50\degr\}$
configuration, synthesized with $\tauz=10$ and rotation rates
$\omega=0.25$, 0.50, and 0.75. For clarity, the $\omega=0.50$ curve is
displaced downward by $0.02\,\fluxstar$, and the $\omega=0.75$ curve
by $0.04\,\fluxstar$.}
\label{fig:omega-curves}
\end{figure}

The three panels of Fig.~\ref{fig:dphase}, showing the light-curve
morphologies in the $i$-$\beta$ plane for $\omega=0.25$, 0.5, and
0.75, illustrate some of the effects of varying the rotation rate. The
most obvious trend is the expansion of the 1-m region as $\omega$ is
increased. This comes almost entirely at the expense of the 0-m
region; almost no change is seen in the 2-m region, either in terms of
its extent, or in terms of the \dphase\ values shown. Again, the
reasons for such behaviour are explored further in \S\ref{sec:torus}.

Focusing now on an individual case, Fig.~\ref{fig:omega-curves} plots
the light curves for a $\ibeta = \{70\degr,50\degr\}$ configuration
[chosen as half-way between the cases shown in panels (ii) and (iii)
of Fig.~\ref{fig:maps}], synthesized for the same three $\omega$
values above. As already noted for Fig.~\ref{fig:dphase}, the rotation
rate has only a minor impact on the phase difference between the light
minima; the measured differences are $\dphase=118\degr$ for
$\omega=0.25$, $\dphase=116\degr$ for $\omega=0.50$, and
$\dphase=124\degr$ for $\omega=0.75$, amounting to only a $\sim 10\%$
variation from a factor-three increase in $\omega$. 

However, a significant increase occurs in the \emph{width} of the
individual light minima toward more-rapid rotation. This indicates
that the magnetospheric plasma is taking relatively longer to transit
across the star, and follows from the approximate scaling of the RRM
inner-edge radius with the Kepler radius,
\begin{equation} \label{eqn:kepler}
\Rkep = \frac{3\, \Rpole}{2\, \omega^{2/3}}
\end{equation}
(cf. TO-05, eqn.~12). As $\omega$ increases, both \Rkep\ and the inner
edge moves in toward the star. Thus, the plasma takes a larger
fraction of a rotation cycle to transit, and broader light minima
ensue.

An additional effect of rapid rotation, seen in the $\omega=0.75$
panel of Fig.~\ref{fig:omega-curves}, is to produce cusping at the
minima of the light curve. This effect, which is the characteristic
hallmark of an optically thick yet geometrically thin disk, arises
because the scale height \hm\ of the disk-like magnetosphere varies as
$\sim \Rkep^{3/2} \sim \omega^{-1}$ (cf. TO-05, eqn.~29).

\subsection{Varying \tauz} \label{ssec:explore-taumag}

\begin{figure}
\begin{center}
\includegraphics{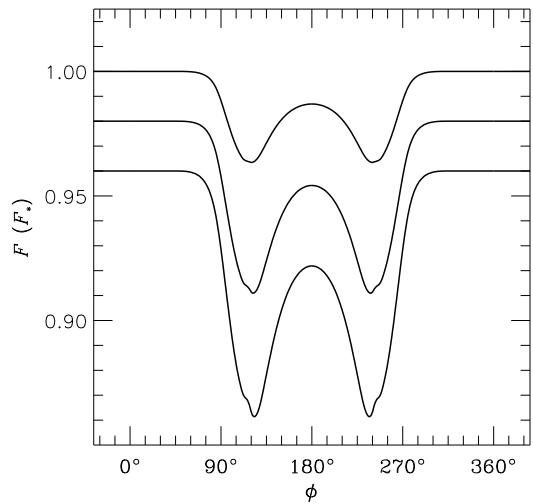}
\end{center}
\caption{Light curves for a $\ibeta = \{70\degr,50\degr\}$
configuration, synthesized for $\omega = 0.50$ and optical depth
scales $\tauz=5$, 10, and 15. For clarity, the $\tauz=10$ curve is
displaced downward by $0.02\,\fluxstar$, and the $\tauz=15$ curve by
$0.04\,\fluxstar$.}
\label{fig:tau-curves}
\end{figure}

For the sake of completeness Fig.~\ref{fig:tau-curves} plots the light
curves for the same $\ibeta = \{70\degr,50\degr\}$ configuration
considered previously, but synthesized this time with a fixed rotation
rate $\omega=0.50$ and varying optical depth scale: $\tauz=5$, 10 and
15. The maximum depths of the light curves -- $0.037\,\fluxstar$,
$0.069\,\fluxstar$ and $0.099\,\fluxstar$, respectively -- increase
approximately linearly with \tauz. This result is a natural
consequence of the fact that the magnetospheric densities (cf. TO-05,
eqn.~35) are proportional to \tauz. As in the preceding section,
however, the change in phase difference is small: in the direction of
increasing \tauz, $\dphase = 120\degr$, $116\degr$ and $114\degr$.

\section{A torus model} \label{sec:torus}

\begin{figure}
\begin{center}
\includegraphics{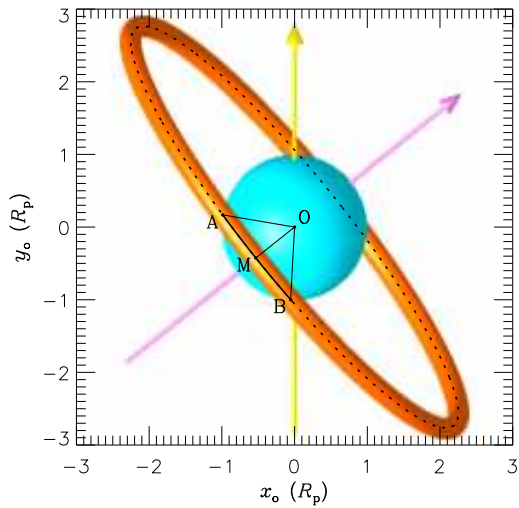}
\end{center}
\caption{The torus model for a $\{i,\angtor,\Rtor\} =
  \{70\degr,50\degr,3.5\,\Rpole\}$ configuration, seen at rotational
  phase $\phase = 92\degr$; the torus is colored orange,
    while the star and axes are as in Fig.~\ref{fig:maps}. The dotted
  line shows the ellipse corresponding to the sky projection of the
  torus; this ellipse is centered at point O, and intersects the
  stellar limb at points A and B. The arc AB has length \arclen, and
  the line OM -- where M is the midpoint of the arc -- is the
  semi-minor axis \sma\ of the ellipse.} \label{fig:torus}
\end{figure}

The previous sections explore how RRM light-curve morphology is
sensitive to parameters such as $i$, $\beta$, and $\omega$. A number
of interesting dependencies emerge, as seen for instance in
Fig.~\ref{fig:dphase}. To gain a measure of insight into these
dependencies, the present section introduces an idealized, analytic
torus model for the formation of the light curves.

In this model, flux variations arise due to obscuration of a spherical
star by an optically thick, oblique, co-rotating torus of
circumstellar material. Assuming that the torus is narrow compared to
the stellar radius, the observed flux at any instant can be
approximated by considering the fraction of the stellar disk obscured:
\begin{equation} \label{eqn:torus-flux}
\flux = \fluxstar \left( 1 - \frac{\arclen \dtor}{\pi \Rstar^{2}}
\right).
\end{equation}
Here, \Rstar\ is the stellar radius, \dtor\ ($\ll \Rstar$) is the
torus cross-sectional diameter, and \arclen\ is the length of the
torus arc (if any) covering the stellar disk. For \Rtor\ not too close
to the stellar radius \Rstar, this arc length is approximated by
\begin{equation} \label{eqn:torus-arclen}
\arclen \approx 2 \sqrt{\max(\Rstar^{2} - \sma^{2},0)},
\end{equation}
where \sma\ is the semi-major axis of the ellipse that is the
projection of the torus on the plane of the sky (see
Fig.~\ref{fig:torus}). If the torus plane is tilted by an angle
\angtor\ to the rotation axis, it can be shown that
\begin{equation} \label{eqn:torus-sma}
\sma = \Rtor |\cos i \cos \angtor + \sin i \sin \angtor \cos \phi|,
\end{equation}
where \Rtor\ is the radius of the torus.

Even before light curves are synthesized, these expressions can be
used to demonstrate that the torus model reproduces the same three
light-curve morphologies found previously. For any light variations
whatsoever to occur, \sma\ must be smaller than \Rstar\ at some point
during the rotation cycle. Assuming that $0\degr \le i \le 90\degr$
and $0\degr \le \angtor \le 90\degr$, this is is equivalent to the
requirement
\begin{equation} \label{eqn:torus-0-m}
\cos (i + \angtor) < \frac{\Rstar}{\Rtor}.
\end{equation}
If this condition is fulfilled, then the light curve exhibits a single
or double minimum morphology depending on whether the torus passes
across the centre of the stellar disk:
\begin{equation} \label{eqn:torus-1-2-m}
\cos (i + \angtor)
\begin{cases}
> 0 & \qquad \mbox{1-m} \\
< 0 & \qquad \mbox{2-m}
\end{cases}
\end{equation}
In the 1-m case, the flux minimum always occurs at phase $\phase =
180\degr$. In the 2-m case, the minima arise when
\begin{equation}
\cos i \cos \angtor = - \sin i \sin \angtor \cos \phi,
\end{equation}
so that the phase difference is given by
\begin{equation} \label{eqn:torus-dphase}
\dphase = 2 \cos^{-1} (\cot i \cot \angtor)
\end{equation}
A similar expression has been derived by \citet[][their
eqn.~2]{ShoBro1990}, but in their case $\beta$ appears in the place of
\angtor.

\begin{figure*}
\begin{center}
\includegraphics{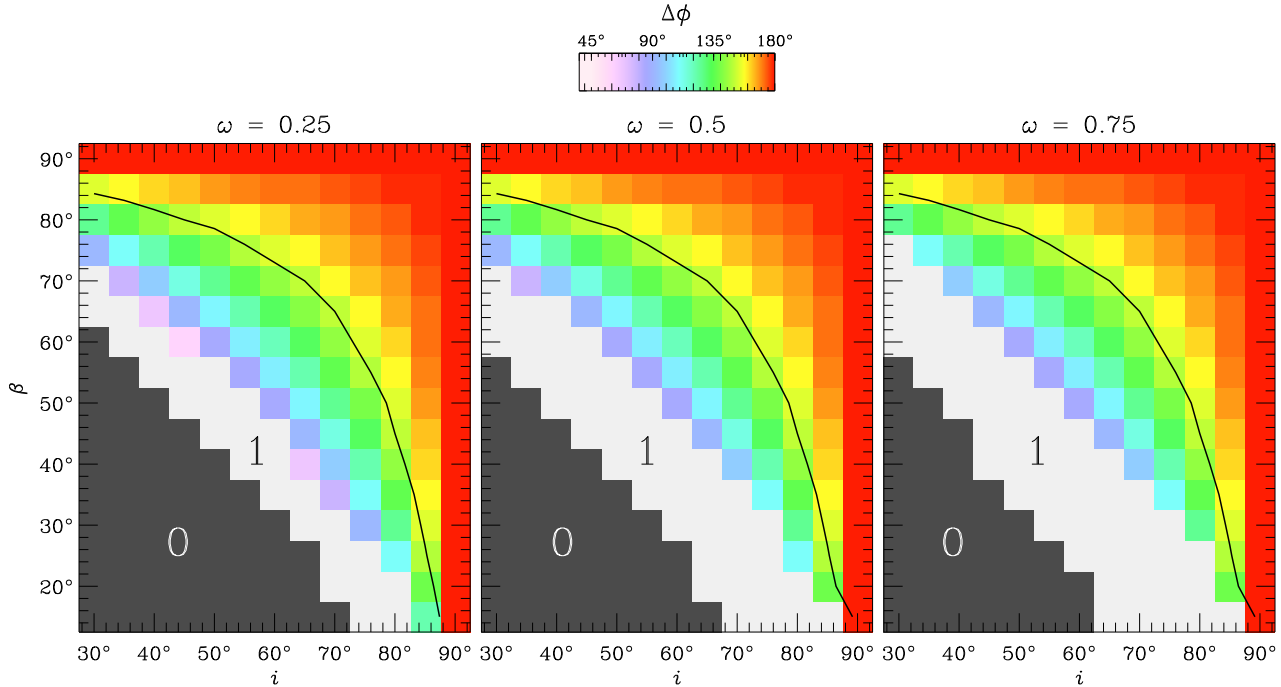}
\end{center}
\caption{As for Fig.~\ref{fig:dphase}, except that the light
curves are calculated using the torus model rather than the RRM
model; in every case, $\dtor = 0.1\,\Rstar$.}
\label{fig:dphase-torus}
\end{figure*}

To now relate the foregoing analysis back to the RRM model,
representative values are chosen for the torus radius \Rtor\ and tilt
angle \angtor. For the former, the Kepler radius \Rkep\
(cf. eqn.~\ref{eqn:kepler}) is the appropriate choice; for the latter,
Appendix~(\ref{app:tilt}) argues that
\begin{equation}
\ang = \beta - \tan^{-1} \left( \frac{\sin 2\beta}{5 + \cos 2\beta}
\right)
\end{equation}
furnishes a good characterization of the mean tilt of the
magnetospheric disk. Adopting these values, and further assuming a
diameter $\dtor = 0.1\Rstar$,
eqns.~(\ref{eqn:torus-flux}--\ref{eqn:torus-sma}) are used to
synthesized a $\beta-i-\omega$ grid of light
curves. These data are analyzed using the algorithm described in
\S\ref{ssec:explore-beta-i}, and Fig.~\ref{fig:dphase-torus} plots the
results in the same format as Fig.~\ref{fig:dphase}.

The two figures shown a good measure of qualitative
agreement. Specific characteristics of the RRM model that are
correctly reproduced by the torus model include:

\begin{itemize}
\item the three differing light-curve morphologies (0-m, 1-m and 2-m),
  and their general locations in the \ibeta\ plane;
\item the fact that the phase differences of 2-m light curves are
  insensitive to rotation rate; this follows from
  eqn.~(\ref{eqn:torus-dphase}), which has no dependence (explicit or
  implicit) on $\omega$;
\item likewise, the fact that the boundary between the 1-m and 2-m
  regions is insensitive to the rotation rate; 
  this follows from eqn.~(\ref{eqn:torus-1-2-m}),
  which again has no dependence on $\omega$;
\item conversely, the fact that the boundary between the 0-m and 1-m
  regions \emph{does} depend on the rotation rate, moving toward lower
  $i$ as $\omega$ is increased; this is due to the identification $\Rtor =
  \Rkep$, which introduces an implicit dependence on $\omega$ into
  eqn.~(\ref{eqn:torus-0-m});
\item finally, although not shown in the figures, the increase in the
  width of 2-m light minima toward more-rapid rotation.
\end{itemize}

At a \emph{quantitative} level, the agreement between the two models
is more limited; for instance, at large $\beta$ the torus model
predicts an extension of the 2-m region to low inclinations, which is
not seen in the RRM model. [In this particular case, the discrepancy
arises because high-$\beta$ magnetospheres are punctured by cone-shape
voids over the rotational poles (see TO-05, fig.~3), and therefore do
not produce light variations when viewed from low inclinations.] Such
discrepancies mean that the torus model should not be used as the sole
basis for data analysis. However, the bulleted list above is ample
proof that -- as a tool for data \emph{interpretation} -- the model
has many useful insights to offer.

\section{Discussion} \label{sec:discuss}

\subsection{Photometric diagnostics} \label{ssec:discuss-diag}

The preceding sections demonstrate that the light variations arising
from a RRM depend on fundamental stellar parameters in a way that is
simple, regular and straightforward to understand. Given these
qualities, it is reasonable to conjecture that observations of these
variations, in He-strong stars and similar objects, could be used in
tandem with other diagnostics such as magnetic field measurements to
establish constraints on the parameters of these stars.

The success of this endeavour rests to a large part on the validity of
the RRM model, since it is from diagrams such as Fig.~\ref{fig:dphase}
-- constructed with the aid of the model -- that reliable diagnostics
are developed. In this respect, the close correspondence between the
observed and synthesized \Halpha\ variations of \sorie\
\citep{Tow2005} has lent considerable empirical support to the
model. Moreover, recent rigid-field hydrodynamical simulations by
\citet*{Tow2007}, and magnetohydrodynamical simulations by
\citet*{udD2007}, have provided confirmation of the theoretical
underpinnings of the model.

Of equal importance, however, is the problem of disentangling the
light variations arising from photospheric abundance inhomogeneities
on the one hand, and from magnetospheric plasma on the other.  For
certain configurations, these two mechanisms can produce variability
that appears ostensibly the same. For instance, the $\ibeta =
\{70\degr,20\degr\}$ magnetospheric light curve in
Fig.~\ref{fig:beta-i-curves}, showing a single, broad minimum, might
easily be confused with the rotational modulation arising from a
spot-like photospheric inhomogeneity. In situations like these, other
diagnostics such as spectroscopy or polarimetry must be used to lift
the degeneracy and correctly identify which mechanism is responsible
for the observed variability.

In the majority of cases, however, the morphologies of the light
curves produced by the alternative mechanisms are rather
different. Especially toward large $i$ and/or $\beta$, the light
variations caused by magnetospheric plasma tend to be much more rapid
(relative to the rotation period) than could plausibly be achieved by
photospheric inhomogeneities. Thus, as a simple rule, faster
variations can be attributed to the magnetosphere, and slower
variations to the photosphere. By way of illustration, the following
section applies this heuristic to five He-strong stars, in some cases
examining whether the constraints provided by Fig.~\ref{fig:dphase}
are consistent with literature values for these stars' parameters, and
in other cases using such constraints to make predictions.




\subsection{Application to He-strong stars} \label{ssec:discuss-app}

\subsubsection{HD~37017}

The light variations of this star \citep[e.g.,][]{Bol1998} show a
smooth, sinusoidal character, favouring a photospheric rather than
magnetospheric origin. \citet{Boh1987} derive $42\degr \lesssim \beta
\lesssim 59\degr$ and $23\degr \lesssim i \lesssim 37\degr$ from
magnetic and period measurements; with reference to
Fig.~\ref{fig:dphase}, these values mostly fall in the 0-m region and
are therefore consistent with the apparent lack of a magnetospheric
signature.

\subsubsection{HD~37479 (\sorie)} \label{sssec:discuss-app-sorie}

As already discussed in \S\ref{sec:intro}, this star exhibits
distinctive 2-m light variations that are consistent with a
magnetospheric origin. That said, certain aspects of its light curve,
such as the unequal depths of the minima and the `emission' feature
after the secondary minimum, appear more likely be of photospheric
origin (see \citealp{Tow2005}; note that these authors explained the
depth asymmetry by instead invoking a decentered dipole).

Nevertheless, this photospheric signal is unlikely to have much effect
on the measured phase difference $\dphase \approx 150\degr$ between
the two minima. This value is indicated in each panel of
Fig.~\ref{fig:dphase} by the contour curves. Along these contours the
combined angle $\beta + i$ varies between $\sim 105\degr$ and $\sim
135\degr$, which led \citet{Tow2005} to adopt the approximate relation
$\beta + i \approx 130\degr$ in developing their model for the
star. This constraint is consistent with the parameters $\ibeta
\approx \{70\degr,56\degr\}$ derived by \citet{ShoBro1990} from
consideration of magnetic and \emph{IUE} data.

\subsubsection{HD~37776}

This star has a rather non-sinusoidal light curve, with a single, very
broad minimum suggestive of a photospheric origin. Detailed modeling
by \citet{Krt2007} demonstrates that, indeed, the photospheric
abundance distribution determined spectroscopically by \citet{Kho2000}
can also faithfully reproduce the star's light curve. Note that the
star has a quadrapolar field topology \citep{ThoLan1985}; although the
RRM model can in principle be applied to any arbitrary field topology,
studies so far (including the present paper) focus on the simplest and
commonest case of a dipole field. Thus, the absence of a clear
magnetospheric signature in the light curve of HD~37776 cannot,
without further modeling, be used to constrain $i$ or $\beta$.

\subsubsection{HD~64740}

This star is the brightest in the He-strong class, yet it exhibits
almost no photometric variations \citep{PedTho1977}. \citet{Boh1987}
derive $\beta \lesssim 76\degr$ and $i \gtrsim 41\degr$ from magnetic
and period measurements. Given the absence of any variability, the
boundary between the 0-m and 1-m regions in Fig.~\ref{fig:dphase}
(assuming $\omega = 0.5$) can be used to further constrain the
parameters as $\beta + 2 i \lesssim 150\degr$.

\subsubsection{HD~182180}

This star has recently been shown to be He-strong, and to exhibit
\Halpha\ and photometric variations, by \citet{Riv2007}. Observational
data remain scarce, but the 2-m, $\dphase = 180\degr$ morphology of
the double-wave light curve is consistent with $\beta = 90\degr$
and/or $i = 90\degr$. The light minima are rather too broad to
convincingly rule out a photospheric origin, but given the star's
extreme rotation period of $0.521\,{\rm d}$, the widths of the minima
are wholly consistent with a near-critical, $\omega \approx 1$
configuration.

\section{Summary} \label{sec:summary}

The RRM model has been used to explore how the light variations due to
magnetospheric plasma depend on the parameters $\beta$, $i$, $\omega$,
and $\tauz$ (\S\ref{sec:explore}). Double-minimum morphologies are
found toward large values of the inclination $i$ and/or obliquity
$\beta$; single-minimum and non-varying morphologies occur for
progressively smaller values of these parameters. In the
double-minimum cases, the phase difference \dphase\ between the minima
depends primarily on $i$ and $\beta$, with almost no sensitivity to
the rotation rate $\omega$ or optical depth scale \tauz.

Insight into these findings has been developed with the aid of an
analytic torus model (\S\ref{sec:torus}). Although not in complete
quantitative agreement with the RRM model, the torus model is able to
explain all of the qualitative results from the parameter-space
exploration. Applying these results, constraints on the obliquity and
inclination of four He-strong stars are derived from their light
curves (\S\ref{sec:discuss}); these are found to be generally
consistent with values quoted in the literature.


\section*{Acknowledgments}

I acknowledge support from NASA \emph{Long Term Space Astrophysics}
grant NNG05GC36G. Interesting conversations with Stan Owocki, John
Landstreet, Tom Bolton and Myron Smith provided much of the motivation
to follow this line of research.


\bibliography{fingerprints}

\bibliographystyle{mn2e}


\appendix

\section{The disk tilt angle} \label{app:tilt}

As discussed in \S\ref{sec:intro}, the RRM model predicts
magnetospheres that resemble a warped disk. A characteristic disk tilt
angle \ang\ with respect to the rotation axis can be obtained by
considering the locus formed by the minima of the dimensionless
effective potential,
\begin{multline} \label{eqn:dpot}
\dpot = -\frac{1}{\drmag\sin^{2}\tmag} - \frac{\drmag^{2}
\sin^{4}\tmag}{2} \left[ \sin^{2}\tmag\,\sin^{2}\pmag + \mbox{}
\phantom{\left( \sin\beta\,\cos\tmag + \cos\beta\,\sin\tmag\,\cos\pmag
\right)^{2}}
  \right. \\
\left. \left( \sin\beta\,\cos\tmag + \cos\beta\,\sin\tmag\,\cos\pmag
\right)^{2} \right]
\end{multline}
(cf. TO-05, eqn.~22). Here, \tmag\ is the magnetic colatitude on
the dipole field line having summit radius \drmag\ (in units of the
Kepler radius \Rkep; cf. eqn.~\ref{eqn:kepler}) and magnetic azimuth
\pmag\footnote{Not to be confused with the rotational phase
\phase.}. The conditions for a potential minimum are that
\begin{equation}
\label{eqn:extremum}
\frac{\partial \dpot}{\partial \tmag} = 0, \qquad
\frac{\partial^{2}\dpot}{\partial \tmag^{2}} > 0.
\end{equation}
As discussed by TO-05, algebraic solution of these equations for
general $\beta$, \drmag, and \pmag\ is not possible. However, the
$\pmag = 90\degr,270\degr$ meridional planes are exceptions -- for
these, it is straightforward to demonstrate that minima occur at the
magnetic equator, $\tmag = 90\degr$.

Immediately adjacent to these special cases, for instance in the
$\pmag = 90\degr + \epsp$ meridional plane for $|\epsp| \ll 1$, the
continuity of the disk requires that the potential minimum should be
at $\tmag = 90\degr + \epst$, where $|\epst| \ll 1$ also. Substituting
this trial solution into the zero-gradient condition
in~(\ref{eqn:extremum}), and neglecting terms of quadratic- or
higher-order in \epsp\ and \epst, leads to the equation
\begin{equation}
- \frac{\drmag^{2} \sin 2\beta}{2} \,\epsp + \left[ - \frac{2}{\drmag} +
  \frac{\drmag^{2} (5 + \cos 2\beta)}{2} \right] \,\epst = 0.
\end{equation}
For field lines extending far outside the Kepler radius (i.e., $\drmag \gg 1$),
the solution is then found as
\begin{equation}
\epst = \frac{\sin 2\beta}{5 + \cos 2\beta} \,\epsp.
\end{equation}
The local tilt angle \angmag\ of the disk with respect to the magnetic
axis is determined from
\begin{equation}
\tan \angmag = -\frac{\epst}{\epsp} = - \frac{\sin 2\beta}{5 + \cos
  2\beta}.
\end{equation}
Thus, the tilt angle with respect to the \emph{rotational} axis
follows as
\begin{equation}
\ang = \beta + \angmag = \beta - \tan^{-1} \left( \frac{\sin 2\beta}{5 + \cos
  2\beta} \right).
\end{equation}
Strictly speaking, this value cannot be applied across the whole of an
RRM disk, due to the warping that arises when $\beta > 0\degr$.
However, as can be seen from Fig.~3 of TO-05, the disk tilt in the
$\pmag=90\degr,270\degr$ meridional planes furnishes a good
approximation to the mean disk tilt, and therefore is well suited to
characterizing the overall disk orientation.


\label{lastpage}

\end{document}